\documentclass[letterpaper, 10 pt, conference]{ieeeconf}  

\IEEEoverridecommandlockouts                              

\usepackage{graphics} 
\usepackage{amsmath,amssymb,amsfonts,bm}

\newtheorem{defn}{Definition}

\usepackage{graphicx}
\usepackage{textcomp}
\let\labelindent\relax
\usepackage{enumitem}

\usepackage{hyperref}

\usepackage{fancyhdr}	

\title{\LARGE \bf
	Trajectory Tracking Control Design for Autonomous Helicopters with Guaranteed Error Bounds
}

\author{Philipp Schitz$^{1,2}$,  Johann C. Dauer$^{1}$, and Paolo Mercorelli$^{2}$
	\thanks{$^{1}$Institute of Flight Systems, German Aerospace Center (DLR), 38108 Braunschweig, Germany (e-mail: philipp.schitz@dlr.de; johann.dauer@dlr.de).}%
	\thanks{$^{2}$Institute for Production Technology and Systems, Leuphana University of Lueneburg, 21335 Lueneburg, Germany (e-mail: paolo.mercorelli@leuphana.de).}%
}
\begin{document}

	\maketitle
	\thispagestyle{fancy}
	\pagestyle{fancy}

	\begin{abstract}
		This paper presents a systematic framework for computing formally guaranteed trajectory tracking error bounds for autonomous helicopters based on Robust Positive Invariant (RPI) sets. The approach establishes a closed-loop translational error dynamics which is cast into polytopic linear parameter-varying form with bounded additive and state-dependent disturbances. Ellipsoidal RPI sets are computed, yielding explicit position error bounds suitable as certified buffer zones in upper-level trajectory planning. Three controller architectures are compared with respect to the conservatism of their error bounds and tracking performance. Simulation results on a nonlinear helicopter model demonstrate that all architectures satisfy the derived bounds, while highlighting trade-offs between performance and the conservatism of the computed invariant set.
	\end{abstract}

\section{Introduction}
Unmanned helicopters combine vertical take-off and landing capabilities with comparatively high payload capacity and range, making them attractive for logistics, inspection, and offshore operations \cite{donkelsAdvancesIntegrationTransport2025}. In such missions, trajectory tracking is typically embedded in a hierarchical architecture: high-level planners generate collision-free reference trajectories, while lower-level controllers ensure their execution.

In most practical systems, uncertainty in trajectory tracking is handled through heuristic safety margins \cite{richterPolynomialTrajectoryPlanning2016,bashirObstacleAvoidanceApproach2023}. While computationally convenient, such approaches lack formal guarantees. If the margin is chosen too small, disturbances may lead to constraint violations. If chosen too large, the planner becomes overly conservative, reducing feasible corridors and mission efficiency. The absence of a systematic link between closed-loop tracking dynamics and trajectory planning remains a central limitation in many frameworks. 

One method to overcome this limitation is the use of invariant sets \cite{blanchiniSetInvarianceControl1999}, which guarantee that a system, once within such a set, will never leave it. However, computation of invariant sets remains difficult for nonlinear systems. A common strategy is thus to bring the system into a linear form and conservatively bound the modeling errors. 
The authors in \cite{greiffInvariantSetPlanning2025} use a motion planning paradigm where nodes in a graph are connected through invariant sets. These sets are generated from a reduced closed-loop model by bounding the errors induced by the attitude dynamics. However, since only position setpoints are considered, the quadrotor is restricted to low velocities. 
Another approach to utilizing robust invariant sets is presented in \cite{althoffOnlineSafetyVerification2015a}, where reachability analysis is employed to validate pre-defined loitering maneuvers for unmanned helicopters. A similar approach is presented in \cite{majumdarFunnelLibrariesRealtime2017}, in which a library of maneuvers with corresponding invariant sets is constructed offline and then used to compose trajectories online. In both cases, the invariant sets are based on linearization along the reference trajectory and are thus valid only for a particular trajectory.
Another line of research is tube-based Model Predictive Control \cite{mayneRobustModelPredictive2005}, where nominal trajectories are planned considering tightened constraints. 
The concept is based on an ancillary controller which keeps the true system close to the reference, with tracking error bounds derived from Robust Positive Invariant (RPI) sets. These error bounds are then used to tighten the constraint sets, ensuring robust constraint satisfaction. This method has been applied to the autonomous landing of unmanned helicopters in our previous works \cite{schitzRobustManeuverPlanning2024} and \cite{schitzRobustHelicopterShip2026}. For helicopters, many existing trajectory tracking architectures \cite{raptisNovelNonlinearBackstepping2011,heModelbasedRealtimeRobust2021,halbeRobustHelicopterSliding2020} produce nonlinear error dynamics and therefore do not lend themselves to RPI set computation. The work in \cite{schitzRobustHelicopterShip2026} overcomes this limitation by introducing a simplified outer-loop model with linear error dynamics. However, the approach only considers constant heading angles and relies on a linearized coupling of acceleration and attitude commands.

\subsection{Contribution and structure of the paper}

In this work, we develop tracking controller architectures to compute RPI sets for unmanned helicopters to be used in any planning framework with sufficiently smooth trajectories. 
Crucially, our proposed method remains valid even for time-varying yaw angles and is adaptable to a wide variety of existing helicopter attitude controllers. 
This is done by replacing the nonlinear coupling of the translation and attitude subsystems by an equivalent acceleration dynamics, resulting a linear outer-loop model. We formally derive disturbance bounds and thus bring the system into a form suitable for computation of RPI sets via semidefinite programming. For outer loop feedback design, we showcase three different architectures and compare the conservatism of their error bounds as well as their tracking performance.


\subsection{Notation}

The identity and zero matrices are denoted as $I$ and $0$, respectively. Given a matrix $P$, we write $P \succ 0 $ ($P \prec 0 $) to denote symmetric positive (negative) definiteness. We interpret the norm operator $|| \cdot ||$ as the (induced) 2-norm. A diagonal matrix $B$ with elements $b_1,\ldots,b_N$ is written as $B = \text{diag}(b_1,\ldots,b_N)$ and the inverse operation  is $(b_1,\ldots,b_N) = \text{diag}^{-1}(B)$. Given vectors or matrices $A_1, \ldots, A_N$, we write the set spanned by their convex hull as $\mathcal{A} = \text{conv}(A_1, \ldots, A_N)$. Given a polytopic set $\mathcal{A}$ we denote its vertices as $\text{vert}(\mathcal{A}) = (A_1, \ldots, A_N)$. 
For a variable $x$, we use $x^{(i)}$ to denote their $i$-th derivative with respect to time and $x^{(i,j)}$ with $i \leq j$ to denote the tupel $(x^{(i)},x^{(i+1)},\ldots,x^{(j)})$.

\section{Problem statement and preliminaries}\label{sec:problem_prelim}

Our aim is to compute explicit position error bounds to be used as buffer zones for collision avoidance in upper-level trajectory planning algorithms. By providing a principled method to arrive at these bounds, the buffer zones are guaranteed to be valid under concrete assumptions that can be monitored during flight. We assume that the planner generates smooth (at least four times smoothly differentiable) three-dimensional position reference trajectories $\bm{p}_r^{(0,4)}$, where $\bm{p}_r = [p_{r,x},p_{r,y},p_{r,z}]^T$. 
We further assume that a at least twice smoothly differentiable heading reference signal is available. This signal may be generated by the planner itself or computed to satisfy a specific objective. In this study, the following constraint on yaw angle of the helicopter is imposed:
\begin{equation}\label{eq:yaw_ref_from_p}
	\psi_r = \text{atan2}(\dot{p}_{r,y},\dot{p}_{r,x}), \quad \dot{p}_{r,x}^2+\dot{p}_{r,y}^2 > 0,
\end{equation}
so that the nose always points into the direction of its velocity vector. Since the necessary derivatives of $\bm{p}_r$ are available, $\dot{\psi}_r$ and $\ddot{\psi}_r$ are readily computed to complete the reference trajectory $\sigma_r = (\bm{p}_r^{(0,4)}, \psi_r^{(0,2)})$. 

For the computation of formally guaranteed error bounds we employ the notion of RPI sets.
\begin{defn}[Robust Positive Invariance]
	Let $\dot{\bm{x}} = f(\bm{x},\bm{d})$ with the state $\bm{x} \in \mathbb{R}^n$ and a disturbance $\bm{d} \in \mathcal{D} \subset \mathbb{R}^p$. A set $\mathcal{Z} \subset \mathbb{R}^n$ is RPI if $\bm{x}(t_0) \in \mathcal{Z}$ implies that $\bm{x}(t) \in \mathcal{Z}$ for all $\bm{d}(t) \in \mathcal{D}$ and $t \geq t_0$.
\end{defn}
Once the state is within the RPI set, we can formally guarantee that it will never leave this set for any disturbance trajectories evolving within $\mathcal{D}$. In this work, we use Linear Matrix Inequality (LMI) tools to compute ellipsoidal RPI sets.
\begin{defn}[Ellipsoidal Set]
	Let $P \in \mathbb{R}^{n\times n}$ be positive definite and $\bm{c} \in \mathbb{R}^n$. An ellipsoidal set with inverse shape matrix $P$ and center $\bm{c}$ is defined as
	\begin{equation}
		\mathcal{E} = \{\bm{x} \in \mathbb{R}^n \;|\; (\bm{x} - \bm{c}) P (\bm{x} - \bm{c}) ^T \leq 1 \}.
	\end{equation}
\end{defn}
In order to obtain a bound on the position error we project the RPI set onto the position subspace. 

For the RPI set computation, we consider polytopic linear systems of the form
\begin{equation}\label{eq:linear_sys}
	\begin{gathered}
		\dot{\bm{x}} = A(\bm{\upsilon}) \bm{x} + E (\Delta C \bm{x} + \bm{d}(t)), \\ A(\bm{\upsilon}) \in \mathcal{A} = \text{conv}(A_1,\ldots,A_N),
	\end{gathered}
\end{equation} 
where $\bm{x} \in \mathbb{R}^n$ denotes the state and $A \in \mathbb{R}^{n \times n}$ is a parameter-varying system matrix with parameter $\bm{\upsilon}$. The possible realizations of the system matrix $A$ are bounded by the convex hull of vertex matrices $A_1,\ldots,A_N$. The disturbances are modeled using state-dependent and additive components. The state dependent disturbance is modeled using a disturbance matrix $\Delta \in \mathbb{R}^{p \times m}$ with bound $||\Delta|| \leq \gamma$ and a matrix $C \in \mathbb{R}^{m \times n}$ to select the relevant outputs from the state vector. The additive disturbance is represented by $\bm{d} \in \mathbb{R}^{p}$ which is bounded by $||\bm{d}|| \leq \bar{d}$. In both cases, $E \in \mathbb{R}^{n \times p}$ maps the disturbances onto the states. 

The ellipsoidal RPI set for system \eqref{eq:linear_sys} is obtained by finding a common Lyapunov function $V = \bm{x}^T P \bm{x}$ for all $A_i \in \mathcal{A}$. Using the Schur complement, we can enforce negativity of $\dot{V}$ with contraction rates $\tau_1, \tau_2 \geq 0$ by ensuring
\begin{equation}\label{eq:LMI_constraint}
	\begin{gathered}
		\begin{bmatrix}			
			M&P E&P E\\
			E^T P&-\tau_1 I&0\\
			E^T P&0&-\tau_2 I
		\end{bmatrix}
		\preceq 0, \quad \forall A_i \in \text{vert}(\mathcal A), \\ 
		M = A_i^T P + P A_i	+ \tau_1 \gamma^2 C^T C	+ \tau_2 \bar{d}^2 P.
	\end{gathered}
\end{equation}
Thus, to minimize the RPI set size, we solve the following optimization problem
\begin{equation}\label{eq:LMI}
	\begin{aligned}
		\min_{P\succ 0, \tau_1, \tau_2 \geq 0} 
		\quad -\log\det(P) \quad 
		\text{s.t.} \quad \eqref{eq:LMI_constraint}
	\end{aligned}
\end{equation}
using semidefinite programming with line search over $\tau_2$. The RPI set is then given by $\mathcal{Z} = \{x\;|\; \bm{x} P \bm{x}^T \leq 1 \}$. Our primary aim in this paper is therefore to design controller architectures that bring the closed-loop error dynamics into the form of \eqref{eq:linear_sys} so that formal tracking error bounds can be established using the optimization problem \eqref{eq:LMI}.

\section{Control oriented helicopter model}\label{sec:model}

We begin by introducing the general form of the helicopter dynamics with state $\bm{x}_H = [\bm{p},\bm{v},\bm{\eta},\bm{\omega},\bm{\chi}]^T$, where $\bm{p} = [p_x, p_y, p_z]^T$, $\bm{v}= [v_x, v_y, v_z]^T$, $\bm{\eta} = [\phi,\theta,\psi]^T$, and $\bm{\omega} = [p, q, r]^T$ denote the inertial position and velocity, the Euler angles and the body-fixed angular rates, respectively. The state $\bm{\chi}$ represents any additional states that describe main or tail rotor dynamics such as blade flapping, inflow or engine behavior. The input $\bm{u}_H = [u_\text{lon},u_\text{lat},u_\text{ped},u_\text{col}]^T$ consists of longitudinal and lateral cyclic, pedal, and collective inputs. The dynamics in their general form are thus given by
\begin{subequations}\label{eq:full_dynamics}
	\begin{align}
		\dot{\bm{p}} &= \bm{v}, \quad
		\dot{\bm{v}} = m^{-1} R {\bm{F}}(\bm{x}_H,\bm{u}_H) + \bm{e}_z g \label{subeq:transl_dynamics}\\
		\dot{\bm{\eta}}& = \Psi \bm{\omega}, \quad 
		\dot{\bm{\omega}} = -(\bm{\omega} \times J{\bm{\omega}}) + {\bm{\tau}}(\bm{x}_H,\bm{u}_H) \label{subeq:att_dynamics} \\
		\dot{\bm{\chi}} &= \bm{f}_\chi(\bm{x}_H,\bm{u}_H) 
	\end{align}
\end{subequations}
where $\Psi$ encodes the kinematic relation between the Euler angles and the angular rates, $\bm{e}_z = [0, 0, 1]^T$, $g$ denotes the gravitational constant, $\bm{F}$ and $\bm{\tau}$ denote general forces and torques, respectively, and $R \in SO(3)$ is the rotation matrix from body-fixed to local inertial coordinates. 

\subsection{Control-oriented model simplification}

As a first step towards a suitable closed-loop model for RPI set computation, we consider a simplified translational helicopter dynamics \eqref{subeq:transl_dynamics}. In particular, we write  
\begin{equation}\label{eq:simple_transl_dynamics}
	\begin{aligned}
		\dot{\bm{v}} = \bm{a} &= -R \bm{e}_z f + \bm{e}_z g + \bar{D} \bm{v}_a + \bm{w}(t), \quad \bar{D} = R D R^T,
	\end{aligned}
\end{equation}
where $f := \alpha u_\text{col}$ with scaling parameter $\alpha$ denotes the mass-normalized thrust acting along the body-fixed $z$-axis, and $D = \text{diag}(d_x,d_y,d_z)$ with $d_x,d_y,d_z \leq 0$ collects linear drag coefficients in body coordinates. The air-relative velocity is defined as $\bm{v}_a := \bm{v} - \bm{v}_W$,
where $\bm{v}_W$ is the estimated mean wind velocity component expressed in the geodetic frame. The term $\bm{w}(t)$ with $||\bm{w}||\leq w_{\max}$ captures all remaining unmodeled effects, including states $\bm{\chi}$, time-varying deviations of the wind from $\bm{v}_W$, and other external disturbances. 

In practice, helicopter attitude controllers are designed using specialized procedures that account for higher-order internal dynamics $\bm{\chi}$ and aerodynamic couplings. Such controllers are typically optimized to track specified reference models. Thus, for the attitude subsystem, we propose a generic interface formulated as reference dynamics for roll $\phi$, pitch $\theta$ and the body frame yaw rate $r$. In particular, for $\bm{\rho} = [\phi_r,\theta_r,f]^T$, define 
\begin{equation}\label{eq:attititude_reference_dynamics}
	\begin{aligned}	
		\ddot{\bm{\rho}} = \Omega^2 (\bm{\rho} - \bm{\rho}_{c}) - 2 \Omega \Xi \dot{\bm{\rho}},
	\end{aligned}
\end{equation}
with bandwidths $\Omega = \text{diag}(\omega_\phi, \omega_\theta, \omega_f)$, damping coefficients $\Xi = \text{diag}(\xi_\phi, \xi_\theta, \xi_f)$, command $\bm{\rho}_c = [\phi_c, \theta_c, f_c]$, and reference angles $\phi_r,\theta_r$. While the reference dynamics on $\phi$ and $\theta$ reflect the inner-loop behavior, the imposed second-order dynamics on $f$ are introduced to guarantee bounded derivatives of the thrust command. This property is required to ensure continuity of the input transformation defined below.
For the yaw rate channel, we assume a first-order tracking dynamics:
\begin{equation}\label{eq:yawrate_reference_dynamics}
	\begin{aligned}	
		\dot{r}_r = -\omega_r (r - r_c).
	\end{aligned}
\end{equation}
with bandwidth $\omega_r$ and yaw rate command $r_c$. By explicitly modeling the closed-loop behavior instead of the controller structure, the subsequent RPI-based analysis remains agnostic to the particular inner-loop implementation.

\subsection{Input transformation and disturbance bounding}
We treat the thrust $f$ and reference attitude $\phi_r$, $\theta_r$, $\psi_r$ as the inputs to the translational subsystem. Define the acceleration command
\begin{equation}\label{eq:input_transformation}
	\bm{a}_c := -R_r \bm{e}_z f_c + \bm{e}_z g,
\end{equation}
where $R_c$ is the rotation matrix associated with the attitude command. The vector $\bm{a}_c$ represents the inertial acceleration generated by the commanded thrust direction and magnitude. In order to invert this mapping, we first express the commanded acceleration in the heading-fixed frame,
\begin{equation}
	\bm{a}_c^H := R_\psi^T \bm{a}_c,
\end{equation}
where $R_\psi$ denotes the yaw rotation. The inverse transform $\bm{\rho}_c = \Phi(\bm{a}_c^H)$ is given by
\begin{equation}\label{eq:dynamics_transform}
	\begin{gathered}
		f_c = \sqrt{{a^H_{c,x}}^2+{a^H_{c,y}}^2+(a^H_{c,z}-g)^2}, \\
		\theta_c = \arctan\left(\frac{a^H_{c,x}}{a^H_{c,z}-g}\right), \quad
		\phi_c = \arcsin\left(\frac{a^H_{c,y}}{f_c}\right).
	\end{gathered}
\end{equation}
Let $\tilde{R} := R R_r^T$ denote the attitude configuration error between the reference attitude $R_r$ and the true attitude $R$. Using the commanded acceleration $\bm{a}_c$ defined in \eqref{eq:input_transformation}, we can rearrange the translational dynamics \eqref{eq:simple_transl_dynamics} into 
\begin{equation}\label{eq:linear_transl_dynamics}
	\begin{aligned}
		\dot{\bm{v}} = \bm{a}_c + \bar{D} \bm{v}_a - (\tilde{R}-I) (g \bm{e}_z -\bm{a}_c)  + \bm{w}(t)
	\end{aligned}
\end{equation}
Using $g \bm{e}_z -\bm{a}_c = R_r \bm{e}_z f$, we obtain
\begin{equation}
	\begin{aligned}
		\dot{\bm{v}} = \bm{a}_c + \bar{D} \bm{v}_{a} - (\tilde{R}-I) R_r \bm{e}_z f + \bm{w}(t).
	\end{aligned}
\end{equation}
To separate nominal and disturbance contributions, introduce a reference velocity $\bm{v}_r$ and define 
$\bm{v}_{a,r} = \bm{v}_r - \bm{v}_W$ and $\bm{e}_v = \bm{v} - \bm{v}_r$. Then $\bm{v}_a = \bm{v}_{a,r} + \bm{e}_v$ and we arrive at
\begin{equation}\label{eq:simplified_translation_dynamics_separated}
	\begin{aligned}
		\dot{\bm{v}} = \bm{a}_c + \bar{D} \bm{v}_{a,r} - (\tilde{R}-I) R_r \bm{e}_z f	+ \bar{D} \bm{e}_v + \bm{w}(t).
	\end{aligned}
\end{equation}
The term $\bar{D} \bm{v}_{a,r}$ will later be compensated via flatness-based feedforward. The remaining terms $(\tilde{R}-I) R_r \bm{e}_z f$ and $\bar{D} \bm{e}_v$ are discussed in the following.

\paragraph{Attitude-induced disturbance} 
The term $(\tilde{R}-I) R_r \bm{e}_z f$ captures the error induced by the thrust misalignment caused by the attitude error. Using $\tilde{R} = RR_r^T$, we obtain
\begin{equation}
	\begin{aligned}
		(\tilde{R} R_r - R_r) \bm{e}_z = (R - R_r) \bm{e}_z.
	\end{aligned}
\end{equation}
Define $R_z := R \bm{e}_z$ and $R_{z,r} := R_r \bm{e}_z$. Both vectors lie on the unit sphere. Let $\delta$ be the angle between them. Then 
\begin{equation}
	|| R_z - R_{z,r} ||^2 = 2(1-\cos(\delta)) \leq 2(1-\cos(\delta_{\max}))
\end{equation}
for a given bounded attitude error $\delta \leq \delta_{\max}$.
Hence, given a thrust bound $|f| \leq f_{\max}$, we obtain a uniform bound
\begin{equation}
	\begin{aligned}
		||(\tilde{R}-I) R_r \bm{e}_z f|| &\leq ||(\tilde{R}-I) R_r \bm{e}_z|| \, |f| \\
		& \leq \sqrt{2(1-\cos(\delta_{\max}))} f_{\max}.
	\end{aligned}
\end{equation}
Combining the attitude-induced error with the disturbance $\bm{d}(t)$ from the simplified translational dynamics \eqref{eq:simple_transl_dynamics}, we obtain
\begin{equation}
	\begin{aligned}
		\bar{d} = \sqrt{2(1-\cos(\delta_{\max}))} f_{\max} + w_{\max},
	\end{aligned}
\end{equation}
which we can directly incorporate into \eqref{eq:LMI}. In practice, $\delta_{\max}$, $f_{\max}$ and  $w_{\max}$ can be estimated from previous flight test or simulation data and should be monitored during flight to ensure the validity of the computed bounds.

\paragraph{Drag-induced disturbance}

Consider now the term $\bar{D} \bm{e}_v$  which represents the velocity-error-dependent disturbance.  Since $R$ is orthonormal, $\bar{D} = RDR^T$ represents a similarity transform. Therefore, the eigenvalues of $\bar{D}$ are identical to those of $D$ and are given by $d_x, d_y, d_z < 0$. Define 
\begin{equation}
	d_{\max} = \max_{i \in \{x,y,z\}} (d_i), \quad d_{\min} = \min_{i \in \{x,y,z\}} (d_i).
\end{equation}
We separate $\bar{D}$ into a the worst-case damping and a residual:
\begin{equation}
	\begin{aligned}
		\bar{D} \bm{e}_v  = d_{\max} \bm{e}_v + \Delta \bar{D} \bm{e}_v, \quad \Delta \bar{D} = \bar{D} - d_{\max} I.
	\end{aligned}
\end{equation}
Hence, the largest eigenvalue of $\Delta \bar{D}$ is $d_{\max} - d_{\min}$ and
\begin{equation}
	\begin{aligned}
		||\Delta \bar{D}|| \leq \gamma = d_{\max} - d_{\min}.
	\end{aligned}
\end{equation}
The term $d_{\max} \bm{e}_v$ is therefore absorbed into the nominal linear dynamics while we set $\Delta$ in \eqref{eq:LMI} to $\Delta \bar{D}$ with $C$ chosen appropriately to extract the velocity error from the state vector of the error dynamics derived in Section \ref{sec:feedback}.

\subsection{Reference model inversion}
The attitude subsystem was previously modeled as a closed-loop reference dynamics in \eqref{eq:attititude_reference_dynamics}. However, through the nonlinear mapping in \eqref{eq:input_transformation}, the translational dynamics cannot be combined with the attitude dynamics to form a linear dynamics. To obtain a representation more suitable for linear control design, we invert the reference model and replace the inner-loop dynamics by an equivalent acceleration dynamics. The particular structure of this dynamics will be detailed during the feedback control design section.
Let $\bm{a}_d$ be desired geodetic acceleration and $\bm{a}^H_d = R_\psi^T \bm{a}_d$ its heading-fixed representation with derivatives
\begin{equation}
	\begin{aligned}	
		\bm{j}^H_d = \dot{\bm{a}}^H_d &= R_\psi^T \bm{j}_d + \dot{R}_\psi \bm{a}_d, \\
		\bm{s}^H_d = \ddot{\bm{a}}^H_d &= R_\psi^T \bm{s}_d + 2 \dot{R}_\psi \bm{j}_d + \ddot{R}_\psi \bm{a}_d.
	\end{aligned}
\end{equation}
Using the inverse input transformation $\bm{\rho} = \Phi(\bm{a}^H_d)$, we can compute the attitude and thrust commands required to generate $\bm{a}_d$. Substituting $\bm{\rho} = \Phi(\bm{a}_d^H)$ into the reference model \eqref{eq:attititude_reference_dynamics} and solving for $\bm{\rho}_c$ yields
\begin{equation}
	\begin{aligned}	
		\bm{\rho}_{c} = \Phi(\bm{a}_d^H) +  2 \Xi \Omega^{-1} \dot{\Phi}(\bm{a}^H_d) + \Omega^{-2} \ddot{\Phi}(\bm{a}^H_d).
	\end{aligned}
\end{equation}
This expression explicitly inverts the closed-loop attitude dynamics so that the inner loop reproduces the desired acceleration $\bm{a}_d$. A direct consequence is that $\bm{a}_d$ must be at least twice differentiable to guarantee the continuity of $\bm{\rho}$. 

A similar inversion is performed for the yaw channel. We impose $\dot{\psi}_r = \dot{\psi}_d$ for a desired $\psi_d$ by an appropriate feed forward command $r_c$. The transformation from Euler angle rate to body rate is
\begin{equation}\label{eq:yawrate_transform}
	\begin{aligned}	
		r_r = -\sin(\phi_r) \dot{\theta}_r
		+ \cos(\phi_r)\cos(\theta_r) \dot{\psi}_r.
	\end{aligned}
\end{equation}
Differentiating this expression and using the first-order reference body yaw rate dynamics \eqref{eq:yawrate_reference_dynamics} gives
\begin{equation*}
	\begin{gathered}	
		\dot{r}_r =  \; \omega_r (r_c - r_r)
		= g_1 \dot{\psi}_d + g_2 \ddot{\psi}_d + g_3, \\
		g_1 = - \sin(\phi_r)\cos(\theta_r)\dot{\phi}_r	- \cos(\phi_r)\sin(\theta_r)\dot{\theta}_r, \\
		g_2 = \cos(\phi_r)\cos(\theta_r), \quad g_3 = - \cos(\phi_r)\dot{\phi}_r\dot{\theta}_r -\sin(\phi_r)\ddot{\theta}_r.
	\end{gathered}
\end{equation*}
Solving for $r_c$ yields
\begin{equation}
	\begin{gathered}
		r_c = r_r + \omega_r^{-1} (g_1 \dot{\psi}_d + g_2\ddot{\psi}_d + g_3).
	\end{gathered}
\end{equation}
With the above inversions, the closed-loop system reduces to the following outer-loop representation for feedback design:
\begin{equation}\label{eq:control_model}
	\begin{aligned}
		\ddot{\bm{p}} = \bm{a}_d + \bar{D} \bm{v}_{a,r} + \bm{w}(t), \quad \ddot{\bm{a}}_{d} = \bm{s}_{d}, \quad  \ddot{\psi}_r = \ddot{\psi}_d.
	\end{aligned}
\end{equation}  
Hence, the nonlinear coupling with the attitude subsystem has been replaced by an acceleration channel dynamics to be designed in the following.

\section{Control}\label{sec:feedback}

In this section, we design a feedforward acceleration signal that compensates the drag term $\bar{D} \bm{v}_{a,r}$. Afterwards, we introduce three different feedback architectures.

\subsection{Drag compensation using flatness-based feedforward}
The translational dynamics with body-fixed linear drag, together with the attitude subsystem, are differentially flat with respect to the position output, as shown in \cite{faesslerDifferentialFlatnessQuadrotor2018}. We exploit this property to compensate the nominal drag term $\bar{D}\bm{v}_{a,r}$ using a feedforward signal $\bm{a}_t$.
Consider the disturbance-free translational dynamics
\begin{equation}
	\bm{a} = -R_r \bm{e}_z f + \bm{e}_z g + \bar{D} \bm{v}_{a,r}
\end{equation}
and parameterize the reference rotation matrix as $R_r = [\bm{x}_B, \bm{y}_B, \bm{z}_B]$. For ideal tracking, $\bar{D} = R_r D R_r^T$. Hence, we can rewrite the dynamics as
\begin{multline}\label{eq:flatness_force_balance}
	f \bm{z}_B
	- (d_x \bm{x}_B^T \bm{v}_{a,r})\bm{x}_B
	- (d_y \bm{y}_B^T \bm{v}_{a,r})\bm{y}_B \\
	- (d_z \bm{z}_B^T \bm{v}_{a,r})\bm{z}_B
	= g \bm{e}_z - \bm{a}.
\end{multline}
Since $\bm{x}_B$, $\bm{y}_B$, $\bm{z}_B$ are orthonormal, 
left-multiplication by $\bm{x}_B^T$ and $\bm{y}_B^T$ eliminates the other axis-aligned terms and we obtain
\begin{gather*}
	\bm{x}_B^T \bm{\alpha} = 0 , \quad \bm{\alpha} = g \bm{e}_z - \bm{a} + d_x \bm{v}_{a,r}, \\
	\bm{y}_B^T \bm{\beta} = 0, \quad \bm{\beta}  = g \bm{e}_z - \bm{a} + d_y \bm{v}_{a,r}.
\end{gather*}
The heading constraint remains unchanged compared to \cite{faesslerDifferentialFlatnessQuadrotor2018} and is enforced by $\bm{y}_C = [-\sin\psi, \; \cos\psi , \; 0]^T$.
The body axes are then constructed as
\begin{equation*}
	\begin{gathered}
		\bm{x}_B = \frac{\bm{y}_C \times \bm{\alpha}}{\|\bm{y}_C \times \bm{\alpha}\|}, \quad
		\bm{y}_B = \frac{\bm{\beta} \times \bm{x}_B}{\|\bm{\beta} \times \bm{x}_B\|}, \quad
		\bm{z}_B = \bm{x}_B \times \bm{y}_B.
	\end{gathered}
\end{equation*}
The body-fixed angular rates $p$, $q$, and $r$ can be obtained by differentiating \eqref{eq:simple_transl_dynamics} once, projecting the dynamics along the body-axes, and solving a linear system. Similarly, the angular accelerations $\dot{p}$, $\dot{q}$, and $\dot{r}$ are computed by differentiating \eqref{eq:simple_transl_dynamics} twice and solving an analogous system. The detailed derivation follows directly from \cite{faesslerDifferentialFlatnessQuadrotor2018} and is omitted for brevity, with modifications only due to the constant wind term and a different sign convention. Thus, for a given reference trajectory $\sigma_r = (\bm{p}_r^{(0,4)},\psi_r^{(0,2)})$, the reference attitude $R_r$ and its derivatives can be computed algebraically.

We now compute feedforward signals for exact tracking of the reference signal in the absence of disturbances and matching initial state. Let $\bm{a}_r$, $\bm{j}_r$, $\bm{s}_r$ denote the reference acceleration, jerk, and snap associated with $\sigma_r$. To cancel the nominal drag term $\bar{D} \bm{v}_{a,r}$, define the feedforward signals
\begin{subequations}
	\begin{align}	
		\bm{a}_t &= \bm{a}_r - \bar{D} \bm{v}_{a,r}, \\
		\bm{j}_t &= \bm{j}_r - (\bar{D} \bm{a}_r + \dot{\bar{D}} \bm{v}_{a,r}), \\
		\bm{s}_t &= \bm{s}_r - (\bar{D} \bm{j}_r + 2\dot{\bar{D}} \bm{a}_r + \ddot{\bar{D}} \bm{v}_{a,r} ),
	\end{align}
\end{subequations}
with times derivatives of $\bar{D}$ given by  
\begin{equation*}
	\begin{gathered}	
		\dot{\bar{D}} = R_r (S_rD+DS_r^T) R_r^T, \\
		\ddot{\bar{D}} = R_r (S_r^2D+DS_r^2 + 2S_rDS_r^T + \dot{S}_rD+ D\dot{S}_r^T) R^T_r,
	\end{gathered}
\end{equation*}
where $S_r$ is the skew-symmetric matrix associated with the reference angular velocity such that $\dot{R}_r = R_r S_r$. 

\subsection{Feedback control}

We present three control architectures that differ in the coordinate frame used for gain definition and acceleration reference dynamics. The central trade-off is between the tightness for their tracking error bounds due to a simple LTI structure and increased flexibility in the selection of the gains which are aligned with the helicopter lateral-longitudinal dynamics.

\begin{figure*}
	\begin{center}
		\includegraphics[width=0.9\linewidth]{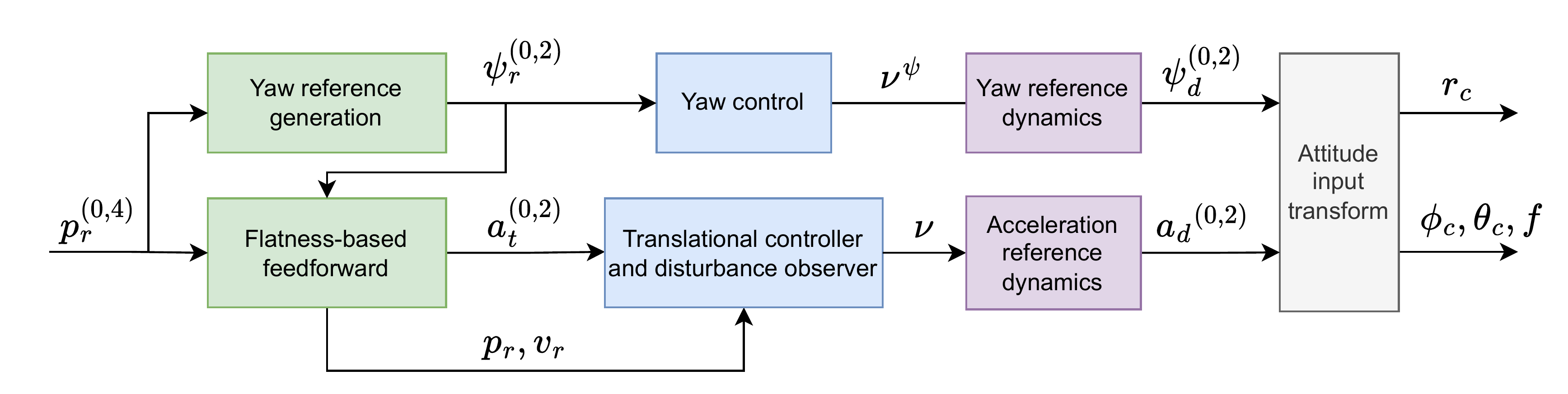} 
		\caption{Generalized overview of the proposed outer loop controller architecture. } 
		\label{fig:control_schematic}
	\end{center}
\end{figure*}

In all cases, we introduce actuator dynamics on the acceleration channel that mimic the bandwidth and damping of the attitude reference model. This avoids unrealistically fast acceleration dynamics, which would result in aggressive attitude commands. The considered architectures are:
\setlist[enumerate]{wide=0pt}
\begin{enumerate}
	\item[\textbf{C-G:}] Geodetic reference model and gains.
	\item[\textbf{C-GH:}] Geodetic reference model with heading-rotated gains.
	\item[\textbf{C-H:}] Heading-fixed reference model and gains.
\end{enumerate}
Figure \ref{fig:control_schematic} shows the complete generalized architecture. Given a reference trajectory $\bm{p}_r$, the yaw reference is generated according to \eqref{eq:yaw_ref_from_p} and is then used for yaw control and the computation of $\bm{a}_t$. The feedforward and its derivatives is then used in the translational controller to generate control inputs $\bm{\nu}$ which then enter the acceleration reference dynamics to be designed. A similar procedure is performed for the yaw channel. Finally, the desired acceleration and yaw commands are transformed into attitude commands going into the inner loop controller. 

\subsection{Disturbance observer}

All architectures employ a nonlinear disturbance observer with linear error dynamics as proposed in \cite{wen-huachenNonlinearDisturbanceObserver2000}. In particular,
\begin{equation} \label{eq:disturbance_estimator}
	\begin{aligned}
		\dot{\bm{z}} &= -L \bm{z} - L \left(\bm{a}_r + g \bm{e}_z + R_r D R_r^T \bm{v}_{a,r} + \bm{\zeta}(\bm{v}) \right) \\
		\hat{\bm{d}} &= \bm{z} + \bm{\zeta}(\bm{v})
	\end{aligned} 
\end{equation}
with an auxiliary variable $\bm{z}$, and the observer gains $L$ and $\bm{\zeta}(\bm{v})$. We select $\bm{\zeta}(\bm{v}) = L \bm{v}$ and $L = \text{diag}(l_1,l_2,l_3)$ with $l_1,l_2,l_3 > 0$. Defining the disturbance estimation error as $\bm{e}_d = \bm{d}(t) - \hat{\bm{d}}$, the disturbance estimation error dynamics become
\begin{equation*}
	\begin{aligned}
		\dot{\bm{e}}_d &= \dot{\bm{d}}(t) + L \bm{z} + L \left(\bm{a}_r + g \bm{e}_z + R_r D R_r ^T \bm{v}_{a,r} + L \bm{v} - \dot{\bm{v}}\right)\\
		&= \dot{\bm{d}}(t) + L (\bm{z} + L \bm{v} - \bm{d}(t)) = \dot{\bm{d}}(t) - L \bm{e}_d,
	\end{aligned} 
\end{equation*}
or, equivalently,
\begin{align}\label{eq:estimator_dynamics}
	\dot{\hat{\bm{d}}} = -L (\hat{\bm{d}} - \bm{d}(t)).
\end{align}
Hence, the disturbance estimate is a low-pass filtered version of the true disturbance. 
We now present each of the controllers considered in this study.

\subsection{C-G: Geodetic gains and reference model}

We first consider an acceleration reference model defined in the geodetic frame:
\begin{equation}
	\begin{gathered}
		\ddot{\bm{a}}_{d} = -\Omega_a^2 (\bm{a}_{d} - \bm{\nu}) - 2 \Omega_a \Xi_a \dot{\bm{a}}_{d} \\ 
		\Omega_a = \text{diag}(\omega_{a,x},\omega_{a,y},\omega_{a,z}),  \quad \Xi_a = \text{diag}(\xi_{a,x},\xi_{a,y},\xi_{a,z})
	\end{gathered}
\end{equation}
Identical bandwidth and damping are chosen in the horizontal directions to preserve symmetry in the $x$-$y$ plane. Define inertial errors $\bm{e}_p := \bm{p} - \bm{p}_r$ and $\bm{e}_v := \bm{v} - \bm{v}_r$. The translational error dynamics become
\begin{equation}
	\begin{aligned}	
		\ddot{\bm{e}}_p &= \dot{\bm{e}}_v = \bm{a}_d - \bm{a}_t + \bm{d}(t)
	\end{aligned}
\end{equation}
Decompose the virtual input as $\bm{\nu} = \bm{\nu}_{ff} + \bm{\nu}_{fb}$ with the feedforward term designed to cancel the filter dynamics:
\begin{equation}
	\begin{gathered}
		\bm{\nu}_{ff} = \bm{a}_t + 2 \Xi_a \Omega_a^{-1} \dot{\bm{a}}_t +  \Omega_a^{-2} \ddot{\bm{a}}_t.
	\end{gathered}
\end{equation}
Let $\bm{e}_a := \bm{a} - \bm{a}_d$. The error dynamics including with the disturbance observer become
\begin{equation}
	\begin{aligned}	
		\ddot{\bm{e}}_p & = \bm{e}_a + \bm{d}(t)\\
		\ddot{\bm{e}}_a & = -\Omega_a^2 (\bm{e}_a - \bm{\nu}_{fb} + K_d \hat{\bm{d}})- 2 \Xi_a \Omega_a \dot{\bm{e}}_a \\
		\dot{\hat{\bm{d}}} & = -L(\hat{\bm{d}} - \bm{d}(t)).
	\end{aligned}
\end{equation}
A simple linear controller is
\begin{equation}
	\bm{\nu}_{fb,1} = - K_p \bm{e}_p - K_v \bm{e}_v - K_a \bm{e}_a - (1 + K_a) \hat{\bm{d}}
\end{equation}
with $K_i = \text{diag}(k_{i,x}, k_{i,y}, k_{i,z})$, $ i \in \{p,v,a\}$. 
The resulting closed-loop system is LTI (no scheduling parameter), however the gains in $x$ and $y$ cannot distinguish between longitudinal and lateral helicopter dynamics.

\subsection{C-GH: Heading-fixed gains, geodetic reference model}

To incorporate distinct longitudinal and lateral behavior, we rotate the gain matrices according to the yaw angle:
\begin{equation}
	K_i^\psi = R_\psi K_i R_\psi^T, \quad i \in \{p,v,a\}
\end{equation}
A rotation of a diagonal matrix with distinct entries $k_{i,x} \neq k_{i,y} $ produces 
a symmetric matrix whose principal axes are rotated by $\psi$. As $\psi$ varies, 
the rotated gain matrices continuously interpolate between two extreme configurations:
\begin{align}
	R_\psi K_i R_\psi^T \Big|_{\psi = 0}
	&= \text{diag}(k_{i,x}, k_{i,y}, k_{i,z})
	= \mathbf{K}_{i,+}, \\
	R_\psi K_i R_\psi^T \Big|_{\psi = \pi/2}
	&= \text{diag}(k_{i,y}, k_{i,x}, k_{i,z})
	= \mathbf{K}_{i,-}.
\end{align}
Hence, the family of heading-dependent gains is contained in the convex hull $\mathbf{K}_i \in \mathrm{conv}(\mathbf{K}_{i,+}, \mathbf{K}_{i,-})$.
The closed-loop system becomes a polytopic LPV system with scheduling parameter $\psi$ and the vertices correspond to the two extreme gain configurations. With this architecture, longitudinal and lateral dynamics can be tuned independently, but the RPI computation must now consider a polytopic system rather than a single LTI model.

\subsection{C-H: Heading-fixed gains and reference model}

The previous architecture defines the acceleration reference model in the geodetic frame. To achieve full alignment with helicopter dynamics, we define both gains and acceleration reference model in the heading-fixed frame:
\begin{gather}
	\ddot{\bm{a}}_{d}^H = -\Omega_a^2 (\bm{a}_{d}^H - \bm{\nu}^H) - 2 \Omega_a \Xi_a \dot{\bm{a}}^H_{d},
\end{gather}
where the superscript $(\cdot)^H$ is used to represent variables in the heading-fixed frame. Define errors $\bm{e}^H_p := R_\psi^T \bm{e}_p$. Using $\dot{R}_\psi = R_\psi S(\dot{\psi})$, where $S(\dot{\psi})^T = - S(\dot{\psi})$, the error dynamics in the heading frame become
\begin{equation}
	\begin{aligned}	
		\dot{\bm{e}}^H_p &= R_\psi^T \bm{e}_v - S(\dot{\psi}) \bm{e}_p^H \\
		\ddot{\bm{e}}^H_p &= \bm{e}_a^H - 2 S(\dot{\psi}) \dot{\bm{e}}^H_p - (S(\dot{\psi})^2 + S(\ddot{\psi})) \bm{e}_p^H + \bm{d}^H(t)\\
	\end{aligned}
\end{equation}
with $\bm{e}_a^H := \bm{a}^H_d - \bm{a}^H_t$ and $\bm{a}^H_t = R_\psi^T \bm{a}_t$. We again design a virtual controller $\bm{\nu}^H = \bm{\nu}^H_{ff} + \bm{\nu}^H_{fb}$ with 
\begin{equation}
	\begin{gathered}
		\bm{\nu}_{ff}^H = \bm{a}_t^H + 2 \Xi_a \Omega_a^{-1} \dot{\bm{a}}^H_t +  \Omega_a^{-2} \ddot{\bm{a}}^H_t.
	\end{gathered}
\end{equation}
Since the control must now also compensate the yaw-induced coupling terms, we take
\begin{multline}
	\bm{\nu}_{fb}^H = -K_p \bm{e}_p^H - K_v \dot{\bm{e}}_p^H - K_a \bm{e}_a^H - (1 + K_a) \hat{\bm{d}}^H \\ + 2 S(\dot{\psi}) \dot{\bm{e}}^H_p + (S(\dot{\psi})^2 + S(\ddot{\psi})) \bm{e}_p^H,
\end{multline}
resulting in the heading-fixed error dynamics
\begin{equation}\label{eq:heading_fixed_error_dynamics}
	\begin{aligned}	
		\ddot{\bm{e}}^H_p & = \bm{e}^H_a + \bm{d}^H(t) - 2 S(\dot{\psi}) \dot{\bm{e}}^H_p - (S(\dot{\psi})^2 + S(\ddot{\psi})) \bm{e}_p^H \\
		\ddot{\bm{e}}_a & = -\Omega_a^2 (\bm{e}_a-\bm{\nu}_{fb}^H)  - 2 \Xi_a \Omega_a \dot{\bm{e}}_a \\
		\dot{\hat{\bm{d}}}^H & = -L(\hat{\bm{d}}^H - \bm{d}^H(t))
	\end{aligned}
\end{equation}
with the disturbance observer designed in the heading frame (derivation omitted for brevity). For bounded yaw rates and accelerations $|\dot{\psi}| \leq \dot{\psi}_{\max}, |\ddot{\psi}| \leq \ddot{\psi}_{\max}$, the system is a polytopic LPV model with vertices obtained by substituting extreme values of $\dot{\psi}$ and $\ddot{\psi}$.
Note that the Coriolis terms containing $S(\psi)$ do not inject or dissipate energy and therefore do not alter the intrinsic stability properties of the system. Nevertheless, they influence transient behavior and attainable performance, and must consequently be included in the RPI analysis.

\subsection{Yaw control}

For all controller architectures, the yaw reference system is chosen as
\begin{gather}\label{eq:yaw_reference_system}
	\ddot{\psi}_d = -\omega_\psi (\dot{\psi}_d - \nu^\psi),
\end{gather}
where $\dot{\psi}_c$ is the yaw rate command and bandwidth chosen as $\omega_\psi = \omega_r$ to match the inner-loop yaw-rate dynamics.
Define $e_{\psi} = \psi - \psi_r$ for $|e_{\psi}| < \pi$. Assuming bounded inner-loop tracking errors, the true yaw rate satisfies $\dot{\psi} = \dot{\psi}_d + \Delta \dot{\psi}$, where $\Delta \dot{\psi}$ is a bounded uncertainty term.
Choose the virtual input
\begin{gather}
	\nu^\psi = \dot{\psi}_r + \omega_\psi^{-1} \ddot{\psi}_r - k_\psi e_\psi, \quad k_\psi > 0
\end{gather}
Substituting into \eqref{eq:yaw_reference_system} and using $\dot{e}_{\psi} = \dot{\psi}_d + \Delta \dot{\psi} - \psi_r$ yields the error dynamics
\begin{equation}
	\begin{aligned}			
		\dot{e}_{\psi} =  \dot{e}_{\psi,d} + \Delta \dot{\psi}, \quad
		\ddot{e}_{\psi,d} = -\omega_\psi (\dot{e}_{\psi,d} + k_\psi e_\psi),
	\end{aligned}
\end{equation}
where $\dot{e}_{\psi,d} = \dot{\psi}_d - \dot{\psi}_r$.
Thus, the yaw subsystem reduces to a stable second-order error dynamics driven only by bounded inner-loop mismatch.

\section{Simulation and discussion}\label{sec:sim_and_discussion}

We compare the three controller architectures in two steps. First, we analyze the size and structure of their RPI sets. Second, we evaluate closed-loop tracking performance on a representative maneuver to assess practical conservatism.

All simulations use the controller gains and acceleration reference model parameters shown in Tab. \ref{tb:parameters}. The inner loop controller is designed to achieve asymptotic damping. Correspondingly, the damping ratio for the acceleration reference models is set to $\xi = 1$. The disturbance estimator gain is $l_1 = l_2 = l_3 = 3$. The disturbance bounds for RPI computation are obtained with $\gamma = 0.4$ and from $\delta_{\max} = 7\, ^\circ$, $f_{\max} = 16\, \text{m/s}^{2}$, and $w_{\max} = 0.5$, resulting in $\bar{d} = 2.5$. For the sake of brevity, we focus our attention only on the horizontal plane, since the vertical dynamics are identical for all architectures.
\begin{table}[hb]
	\begin{center}
		\caption{Controller gains and reference model parameters.}\label{tb:parameters}
		\begin{tabular}{c|c|c|c|c}
			&  $\text{diag}^{-1}(K_p)$ & $\text{diag}^{-1}(K_v)$ &  $\text{diag}^{-1}(K_a)$ & $\text{diag}^{-1}(\Omega)$ \\ \hline
			C-G & $[1,1,2]$ & $[1.4,1.4,3]$ & $[0.5,0.5,1]$ &$[7.5,7.5,12]$ \\
			C-GH &  $[1,1.5,2]$ & $[1.4,2,3]$ & $[0.5,0.5,1]$  &$[7.5,7.5,12]$ \\
			C-H & $[1,1.5,2]$ & $[1.4,2,3]$ & $[0.5,0.5,1]$  & $[7.5,10,12]$ \\
		\end{tabular}
	\end{center}
\end{table}%

\subsection{Error bounds}

\begin{figure}
	\begin{center}
		\includegraphics[width=0.95\linewidth]{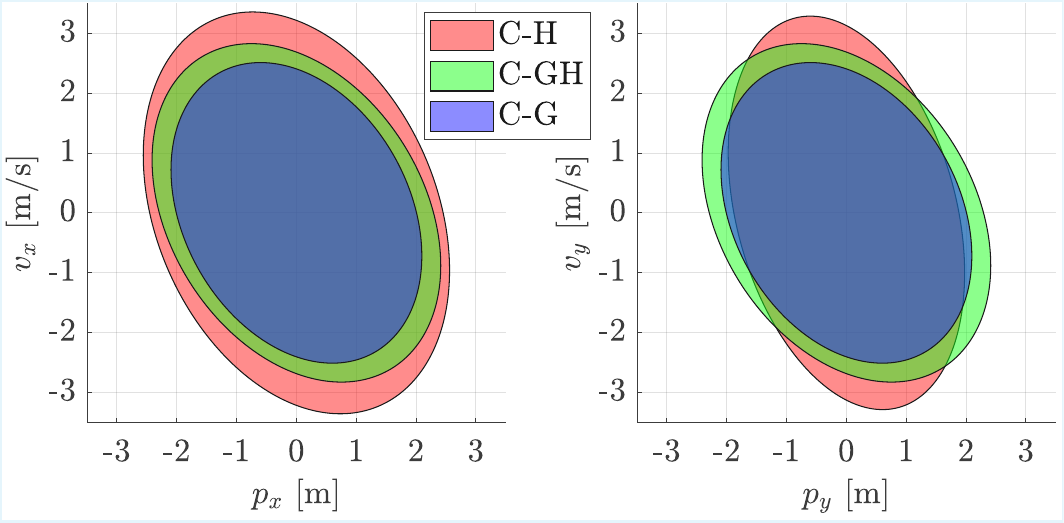} 
		\caption{Projections of RPI sets onto the position-velocity subspace for the three considered feedback controllers.} 
		\label{fig:rpi_sets}
	\end{center}
\end{figure}

The three controllers reflect different trade-offs between dynamical fidelity and conservatism in the LMI-based RPI computation. Figure \ref{fig:rpi_sets} shows the projection of the RPI sets onto the horizontal position–velocity subspace. The LMI optimization is performed once offline and takes roughly two seconds including the line search for all three cases.

\paragraph{C-G} The purely geodetic controller yields the smallest position bound, despite the most conservative gain selection. Its simple closed-loop structure leads to reduced conservatism in the LMI formulation. Since the error is expressed in the geodetic frame, the bounds in the  $x$- and $y$-directions are identical and independent of yaw.

\paragraph{C-GH} Introducing heading-dependent gains slightly enlarges the RPI set. Although lateral gains are more aggressive, the LMI synthesis employs a common Lyapunov function for all admissible gain combinations. This aggregates the worst-case properties of both axes into a single invariant set. The bound remains yaw-invariant.

\paragraph{C-H} 
The heading-frame architecture models the attitude dynamics most accurately and allows axis-specific tuning consistent with roll and pitch behavior. However, the influence of the additional nonlinear Coriolis terms have to be modeled conservatively ($\dot{\psi}_{\max} = \ddot{\psi}_{\max} = 0.5$), thus increasing the RPI set size.
In the slower $x$-axis, this controller produces the largest bound. In contrast, it yields the smallest bound in the $y$-axis, despite the added conservatism. A key structural difference is the loss of yaw invariance: the RPI set rotates with the vehicle heading. This increases complexity in trajectory planning and collision checking, as obstacle intersection tests must account for orientation.

\subsection{Trajectory tracking control}
\begin{figure}
	\begin{center}
		\includegraphics[width=1\linewidth]{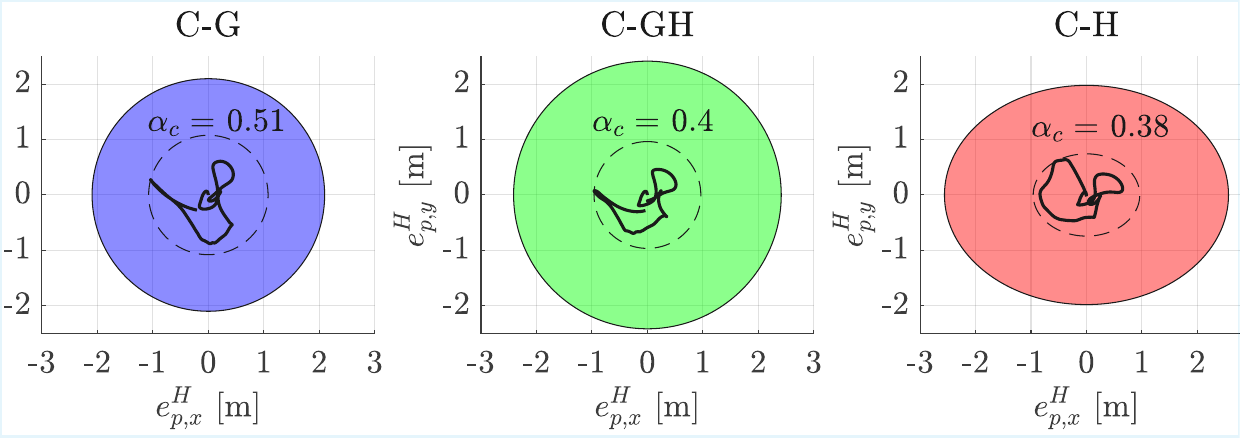} 
		\caption{Projection of RPI sets onto the $x$-$y$ position subspace with heading-fixed error trajectories for all three feedback architectures. The scalar $\alpha_c$ denotes the minimum scaling such that the error trajectory is contained within the ellipsoid with inverse shape matrix $P_c = \alpha_cP$, shown with dashed lines.} 
		\label{fig:RPI_traj}
	\end{center}
\end{figure}

\begin{figure}
	\begin{center}
		\includegraphics[width=0.9\linewidth]{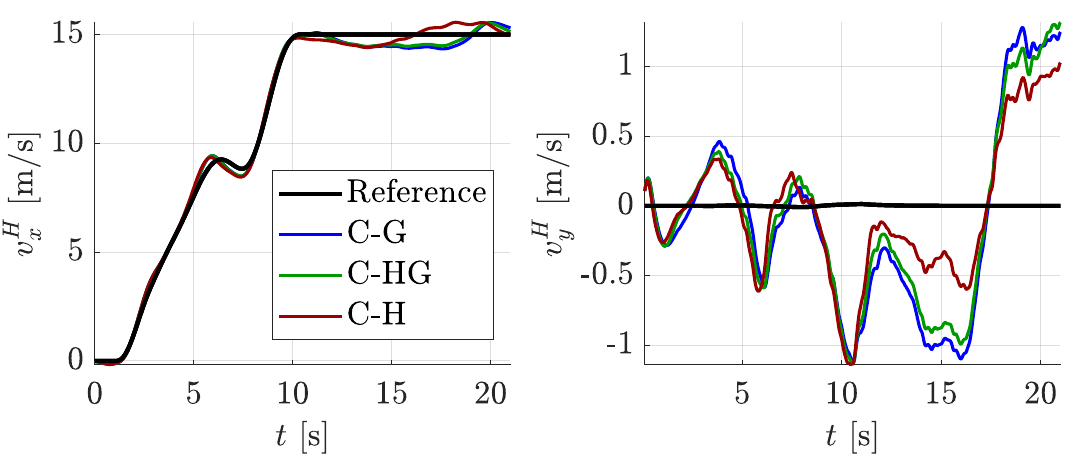} 
		\includegraphics[width=0.9\linewidth]{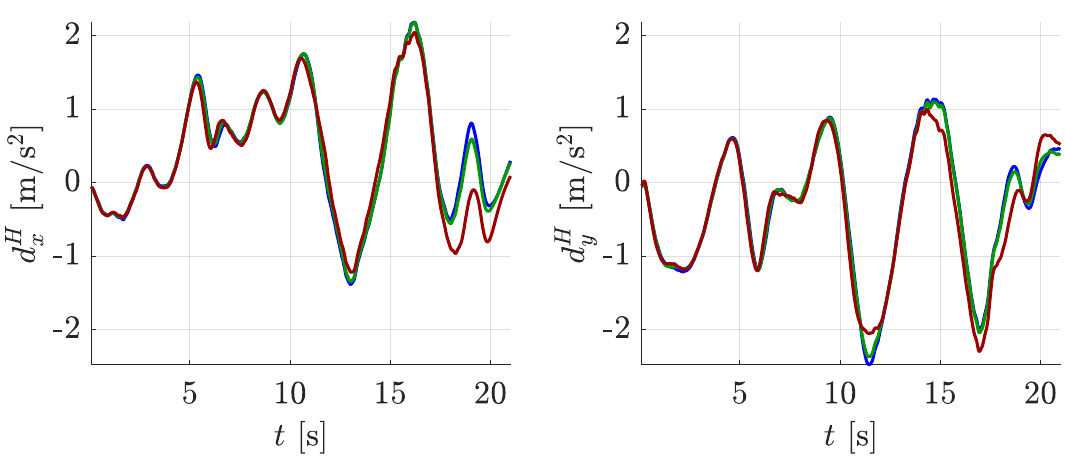} 
		\includegraphics[width=0.9\linewidth]{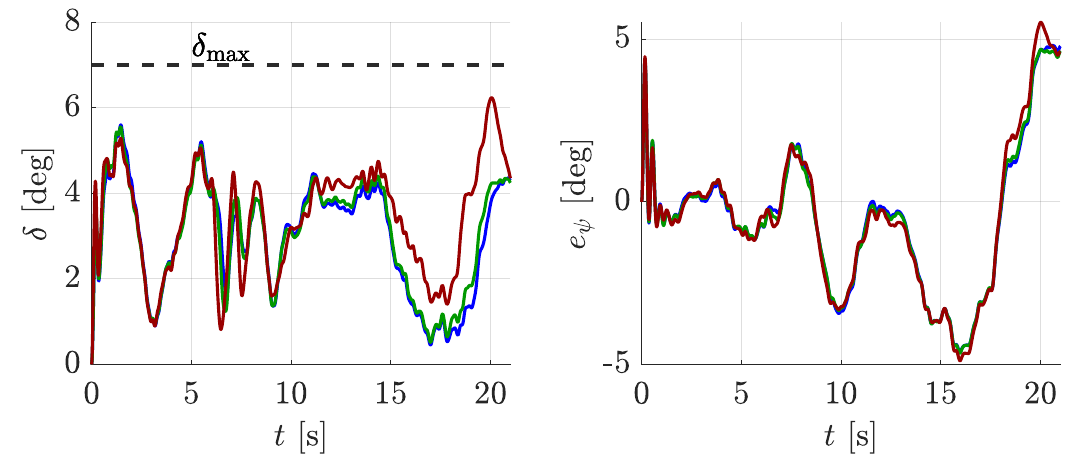} 
		\caption{Trajectories of horizontal velocities in the heading frame (top row), disturbances in the heading frame (middle row), tilt angle error (bottom left), and yaw error (bottom right).} 
		\label{fig:timeseries_plots}
	\end{center}
\end{figure}

Next, we examine the control performance and evaluate conservatism using a fairly aggressive example maneuver. In particular, we consider an acceleration into a loiter with a 30~m radius and at 15~m/s under a mean wind speed of 7~m/s in the direction of the $y$ axis. This results in mean roll angle of approximately 30 degrees during the loiter. The reference trajectory is generated using the minimum-snap framework of \cite{mellingerMinimumSnapTrajectory2011}. The controllers are tested on the full nonlinear model \eqref{eq:full_dynamics} representing the midiARTIS helicopter, a small research helicopter operated by the German Aerospace Center. More information regarding its modeling and the attitude controller can be found in \cite{petitSystemIdentification2025}. 
\begin{figure}
	\begin{center}
		\includegraphics[width=0.9\linewidth]{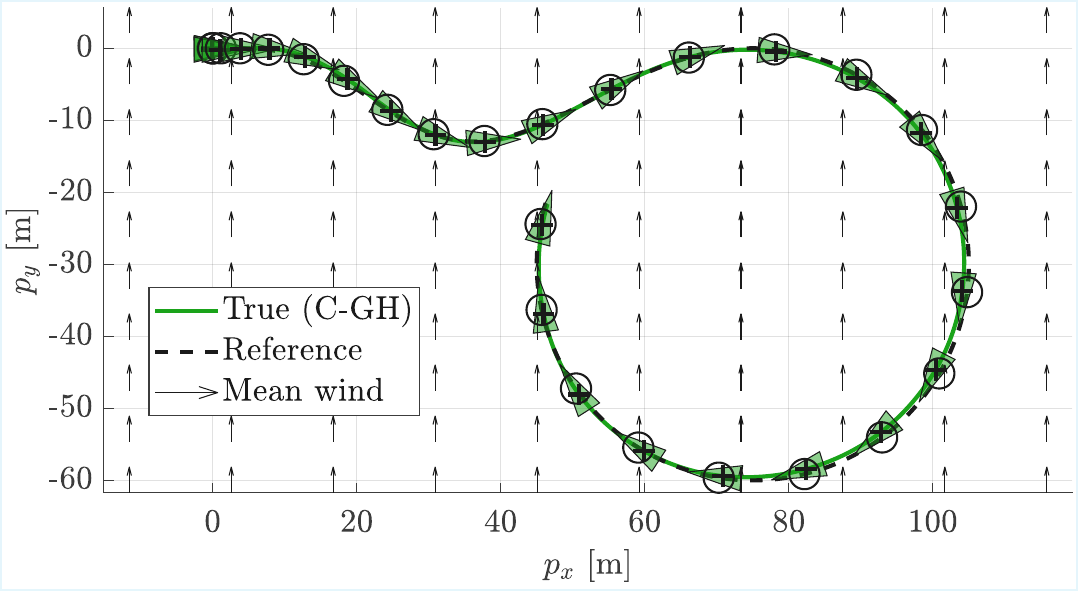} 
		\caption{Position trajectories of the considered maneuver in the $x$-$y$-plane for the C-GH controller with snapshots of the helicopter heading, position and the associated projection of the RPI ellipsoid onto the horizontal position subspace.} 
		\label{fig:xy_circle}
	\end{center}
\end{figure}
Figure \ref{fig:xy_circle} shows the position reference trajectory of the considered maneuver for the controller C-GH along with error bound snapshots. Figure \ref{fig:timeseries_plots} shows the associated time evolution of the horizontal velocity, disturbance estimation, as well as tilt and yaw angle errors. Lastly, Figure \ref{fig:RPI_traj} depicts the position errors and their computed bound for the respective controller architecture.

Notably, as can be seen in Figure \ref{fig:RPI_traj}, all controllers track the trajectory accurately and remain within their computed RPI bounds despite wind disturbances. This confirms that the inversion-based outer-loop design remains valid for the full nonlinear system. 
Performance differences align with the RPI analysis: {C-G} exhibits the largest tracking deviations due to conservative gains, but its RPI bound is comparatively tight relative to the observed error for this specific maneuver (51\% coverage of the computed RPI set). {C-GH} reduces position deviations while maintaining similar qualitative behavior. {C-H} shows tracking position errors comparable to {C-GH}. However, the RPI sets of both {C-GH} and {C-H} are significantly larger than the observed maximum deviation (error trajectories show 40\% and 38\% coverage of the RPI set, respectively), indicating increased conservatism in the robustness analysis.

In Figure \ref{fig:timeseries_plots}, {C-G} and {C-GH} behave similarly, with modest differences in tilt and velocity errors. {C-H} displays the most aggressive lateral response, resulting in larger tilt-angle errors but reduced velocity error in the $y$-direction. Overall, increasing dynamical fidelity improves axis-specific tracking performance but enlarges the certified invariant set and, in the heading-frame case, sacrifices yaw invariance.
Note that the maximum tilt angle and maximum thrust magnitude (not shown) used for $\bar{d}$ are not violated during the maneuver, ensuring the soundness of the theoretical guarantees provided by the bound. The fact that the computed position error bounds remain valid under observable assumptions is one of the key advantages of the proposed method.

\section{Conclusion and future work}\label{sec:conclusion}

This paper presented a framework for computing formally guaranteed trajectory tracking error bounds for unmanned helicopters. By combining flatness-based feedforward design, inversion of closed-loop attitude reference dynamics, and outer-loop acceleration controllers, the nonlinear helicopter model was reduced to a polytopic LPV error system suitable for ellipsoidal RPI set computation via LMIs.
Three feedback architectures were analyzed. The purely geodetic design yields the smallest invariant set, despite conservative gain tuning. Introducing heading-dependent gains improves axis-specific tracking performance but enlarges the RPI set. The fully heading-fixed architecture most accurately reflects helicopter dynamics. However, the additional yaw-induced coupling terms increase conservatism and lead to orientation-dependent bounds. Nonlinear simulations demonstrated that all architectures respect their computed invariant. The proposed framework establishes an explicit link between closed-loop tracking dynamics and motion planning by providing formally guaranteed, monitorable buffer zones. 
However, a central limitation of this study is the simulation of a single scenario. Future work should therefore focus on validating the proposed method across different maneuvers and assessing its applicability in practice via flight tests. On the technical side, tighter invariant set approximations and integration of the bounds into online trajectory optimization and monitoring schemes could be addressed.

\bibliography{bibliography.bib} 

@article{faesslerDifferentialFlatnessQuadrotor2018,
	title = {Differential {{Flatness}} of {{Quadrotor Dynamics Subject}} to {{Rotor Drag}} for {{Accurate Tracking}} of {{High-Speed Trajectories}}},
	author = {Faessler, Matthias and Franchi, Antonio and Scaramuzza, Davide},
	year = {2018},
	journal = {IEEE Robotics and Automation Letters},
	volume = {3},
	number = {2},
	pages = {620--626},
	issn = {2377-3766, 2377-3774},
	doi = {10.1109/LRA.2017.2776353}
}

@article{schitzRobustManeuverPlanning2024,
	title = {Robust {{Maneuver Planning With Scalable Prediction Horizons}}: {{A Move Blocking Approach}}},
	shorttitle = {Robust {{Maneuver Planning With Scalable Prediction Horizons}}},
	author = {Schitz, Philipp and Dauer, Johann C. and Mercorelli, Paolo},
	year = {2024},
	journal = {IEEE Control Systems Letters},
	volume = {8},
	pages = {1907--1912}
}

@unpublished{petitSystemIdentification2025,
	title = {System Identification Driven High-Performance Controller
	Design For An Agile Small Rotorcraft {UAV}},
	author = {Petit, P and Seher-Weiß, S and Wartmann, J and Donkels, A},	
	note = {presented at the 51st European Rotorcraft Forum, Venice, Italy, Sept. 9-12, 2025}
}

@article{wen-huachenNonlinearDisturbanceObserver2000,
	title = {A Nonlinear Disturbance Observer for Robotic Manipulators},
	author = {Chen, Wen-Hua and Ballance, D.J. and Gawthrop, P.J. and O'Reilly, J.},
	year = {2000},
	month = aug,
	journal = {IEEE Transactions on Industrial Electronics},
	volume = {47},
	number = {4},
	pages = {932--938}
}

@article{greiffInvariantSetPlanning2025,
	title = {Invariant {{Set Planning}} for {{Quadrotors}}: {{Design}}, {{Analysis}}, {{Experiments}}},
	shorttitle = {Invariant {{Set Planning}} for {{Quadrotors}}},
	author = {Greiff, Marcus and Sinhmar, Himani and Weiss, Avishai and Berntorp, Karl and Di Cairano, Stefano},
	year = {2025},
	month = mar,
	journal = {IEEE Transactions on Control Systems Technology},
	volume = {33},
	number = {2},
	pages = {449--462},
	issn = {1063-6536, 1558-0865, 2374-0159},
	doi = {10.1109/TCST.2024.3492813}
}

@inproceedings{mellingerMinimumSnapTrajectory2011,
	title = {Minimum Snap Trajectory Generation and Control for Quadrotors},
	booktitle = {2011 {{IEEE International Conference}} on {{Robotics}} and {{Automation}}},
	author = {Mellinger, Daniel and Kumar, Vijay},
	year = {2011},
	month = may,
	pages = {2520--2525},
	publisher = {IEEE},
	address = {Shanghai, China}
}

@article{blanchiniSetInvarianceControl1999,
	title = {Set Invariance in Control},
	author = {Blanchini, F.},
	year = {1999},
	month = nov,
	journal = {Automatica},
	volume = {35},
	number = {11},
	pages = {1747--1767},
	issn = {00051098}
}

@inproceedings{althoffOnlineSafetyVerification2015a,
	title = {Online Safety Verification of Trajectories for Unmanned Flight with Offline Computed Robust Invariant Sets},
	booktitle = {2015 {{IEEE}}/{{RSJ International Conference}} on {{Intelligent Robots}} and {{Systems}} ({{IROS}})},
	author = {Althoff, Daniel and Althoff, Matthias and Scherer, Sebastian},
	year = {2015},
	month = sep,
	pages = {3470--3477},
	publisher = {IEEE},
	address = {Hamburg, Germany}
}

@article{mayneRobustModelPredictive2005,
	title = {Robust Model Predictive Control of Constrained Linear Systems with Bounded Disturbances},
	author = {Mayne, D.Q. and Seron, M.M. and Rakovi{\'c}, S.V.},
	year = {2005},
	month = feb,
	journal = {Automatica},
	volume = {41},
	number = {2},
	pages = {219--224},
	issn = {00051098}
}

@incollection{richterPolynomialTrajectoryPlanning2016,
	title = {Polynomial {{Trajectory Planning}} for {{Aggressive Quadrotor Flight}} in {{Dense Indoor Environments}}},
	booktitle = {Robotics {{Research}}},
	author = {Richter, Charles and Bry, Adam and Roy, Nicholas},
	editor = {Inaba, Masayuki and Corke, Peter},
	year = {2016},
	volume = {114},
	pages = {649--666},
	publisher = {Springer International Publishing}
}

@article{bashirObstacleAvoidanceApproach2023,
	title = {An Obstacle Avoidance Approach for {{UAV}} Path Planning},
	author = {Bashir, Nouman and Boudjit, Saadi and Dauphin, Gabriel and Zeadally, Sherali},
	year = {2023},
	month = dec,
	journal = {Simulation Modelling Practice and Theory},
	volume = {129},
	pages = {102815},
	issn = {1569190X}
}

@article{majumdarFunnelLibrariesRealtime2017,
	title = {Funnel Libraries for Real-Time Robust Feedback Motion Planning},
	author = {Majumdar, Anirudha and Tedrake, Russ},
	year = {2017},
	month = jul,
	journal = {The International Journal of Robotics Research},
	volume = {36},
	number = {8},
	pages = {947--982},
	issn = {0278-3649, 1741-3176}
}

@article{donkelsAdvancesIntegrationTransport2025,
	title = {Advances on the Integration of Transport Drones into Offshore Wind Farms},
	author = {Donkels, Alexander and Cain, Sebastian and Wilhelm, Jannik and Dippmar, John and Heil, Detlef and Janke, Jonas and Bruns, Tilmann},
	year = {2025},
	month = jun,
	journal = {CEAS Aeronautical Journal},
	issn = {1869-5582, 1869-5590}
}

@article{raptisNovelNonlinearBackstepping2011,
	title = {A {{Novel Nonlinear Backstepping Controller Design}} for {{Helicopters Using}} the {{Rotation Matrix}}},
	author = {Raptis, Ioannis A. and Valavanis, Kimon P. and Moreno, Wilfrido A.},
	year = {2011},
	month = mar,
	journal = {IEEE Transactions on Control Systems Technology},
	volume = {19},
	number = {2},
	pages = {465--473},
	issn = {1063-6536, 1558-0865}
}

@article{heModelbasedRealtimeRobust2021,
	title = {Model-Based Real-Time Robust Controller for a Small Helicopter},
	author = {He, Miaolei and He, Jilin and Scherer, Sebastian},
	year = {2021},
	month = jan,
	journal = {Mechanical Systems and Signal Processing},
	volume = {146},
	pages = {107022},
	issn = {08883270}
}

@article{halbeRobustHelicopterSliding2020,
	title = {Robust {{Helicopter Sliding Mode Control}} for {{Enhanced Handling}} and {{Trajectory Following}}},
	author = {Halbe, Omkar and Hajek, Manfred},
	year = {2020},
	month = oct,
	journal = {Journal of Guidance, Control, and Dynamics},
	volume = {43},
	number = {10},
	pages = {1805--1821},
	issn = {1533-3884},
	doi = {10.2514/1.G005183},
	urldate = {2026-02-26},
	langid = {english}
}

@unpublished{schitzRobustHelicopterShip2026,
	title = {Robust {Helicopter Ship Deck Landing With Guaranteed Timing Using Shrinking-Horizon Model Predictive Control}},
	author = {Schitz, Philipp  and Mercorelli, Paolo and Dauer, Johann C.},	
	note = {Accepted to the American Control Conference 2026, New Orleans, Louisiana, May 26-29, 2026. Available: \url{https://arxiv.org/abs/2602.22714}}
}

\end{document}